# Hysteretic magnetoresistance and thermal bistability in a magnetic two-dimensional hole system

Ursula Wurstbauer[1,2‡], Cezary Śliwa[3], Dieter Weiss[1], Tomasz Dietl[3,4] and Werner Wegscheider[1*†]

[1] Institute of Experimental and Applied Physics, University of Regensburg, D-93040 Regensburg, Germany

[2] Institute of Applied Physics, University of Hamburg, D-20355 Hamburg, Germany

[3] Institute of Physics, Polish Academy of Sciences, al. Lotników 32/46, PL-02 668 Warszawa, Poland

[4] Institute of Theoretical Physics, University of Warsaw, ul. Hoża 69, PL 00 681 Warszawa, Poland

*e-mail: werner.wegscheider@phys.ethz.ch

[†]present address: Solid State Physics Laboratory, ETH Zurich, Switzerland

[‡]present address: Department of Physics, Columbia University, USA

**Colossal negative magnetoresistance and the associated field-induced insulator-to-metal transition, the most characteristic features of magnetic semiconductors, are observed in *n*-type rare earth oxides[1] and chalcogenides[2], *p*-type manganites[3], *n*-type[4,5] and *p*-type diluted magnetic semiconductors (DMS)[4,6] as well as in quantum wells of *n*-type DMS[7-9]. Here, we report on magnetostransport studies of Mn modulation-doped InAs quantum wells, which reveal a magnetic field driven and bias voltage dependent insulator-to-metal transition with abrupt and hysteretic changes of resistance over several orders of magnitude. These phenomena coexist with the quantised Hall effect in high magnetic fields. We show that the exchange coupling between a hole and the parent Mn acceptor produces a magnetic anisotropy barrier that shifts the spin relaxation time of the bound hole to a 100 s range in compressively strained quantum wells. This bistability of the individual Mn acceptors explains the hysteretic behaviour while opening prospects for information storing and processing. At high bias voltage another bistability, caused by the overheating of electrons[10], gives rise to abrupt resistance jumps.**

Molecular beam epitaxy has been employed to grow Mn modulation doped compressively-strained InAs quantum wells (QWs) embedded in InAlAs/InGaAs host material with an In mole fraction of 75% (for details, see, ref. 11). In this material system Mn substitutes the group III elements (In, Al or Ga), providing a localised spin of $S = 5/2$ and a hole[6,12,13], in contrast to II-VI materials, where Mn is an isoelectric impurity[4,7-9,14]. Our transport studies have been accomplished on Mn modulation doped InAs quantum well structures depicted schematically in Fig. 1. In the "normal" sample (Fig. 1b) the Mn doping is done after the InAs/InGaAs layer growth, so that the InAs channel is free of Mn[11,15]. In the "inverted" doped structures (Fig. 1a) the Mn doped layer is deposited *prior* to the growth of the 4 nm thick InAs channel which is 7.5 nm spaced from the Mn doped layer. This, leads to a significant concentration of Mn in the InAs channel, estimated by secondary ion mass spectroscopy (SIMS) to be ~1% of the maximum doping concentration[11].

Figure 1 documents remarkable differences in the transport properties at 1.6 K of "normal" and "inverted" doped samples with comparable two-dimensional hole densities of $p = 4.3 \cdot 10^{10}$ cm$^{-2}$ and $p = 4.4 \cdot 10^{10}$ cm$^{-2}$, respectively. Both types of samples show temperature and field dependent resistances typical for modulation-doped structures, including pronounced Shubnikov-de Haas (SdH) oscillations and quantum Hall plateaus in high fields. This clearly demonstrates the two-dimensional nature of the charge carrier system and the absence of parallel conductance, a conclusion consistent with a sufficiently low Mn concentration to prevent an insulator-to-metal transition of Mn acceptors in the InAlAs barrier.

According to the experimental findings summarised in Figs. 1a, 2, and 3 the "inverted" structure (sample A) shows a dramatic and temperature dependent increase of resistance in the zero-field range indicating a strong hole localisation under these conditions. The application of a perpendicular magnetic field leads to a colossal negative magnetoresistance resulting eventually in the quantum Hall insulator transition at $B_T \cong 4$ T, above which the longitudinal resistance *decreases* with lowering temperature (Fig. 2). In this high field range the quantum Hall effect emerges with a well developed plateau and a corresponding zero-resistance state at low temperatures (Figs. 1a and 2).

Because of the extremely high resistance values at low temperatures and magnetic fields, measurements in this region have been performed in a two-terminal geometry by applying a constant voltage $U_{bias}$ to the sample and measuring the current $I$ according to $R = U_{bias}/I$. As depicted in Figs. 3a-3c, field-induced holes' delocalisation is accompanied by resistance jumps over several orders of magnitude from above $10^{11}$ down to below $10^6$ Ω, particularly abrupt for a relatively high bias voltage of $U_{bias} = 0.5$ V (Figs. 3b, 3c), where the dependence $R(U_{bias})$ is non-linear at low temperatures (Fig. 3d). At the same time, the resistance shows evidence of a notable hysteretic behaviour when sweeping the magnetic field across zero, i.e. the resistance at high and low B is asymmetric with respect to zero field. (Figs. 3a-3c). As seen, the changes and jumps of the resistance as well as the hysteretic behaviour diminish and finally disappear above ~0.6 K.

According to results collected in Fig. 3c, hole accumulation by a negative top-gate voltage shifts the system away from the localisation boundary to a region where the resistance is smaller and much less temperature dependent. In this regime, (abrupt) resistance changes vanish and hystereses are reduced. Depletion of the channel moves both $B_{jump}$ and the resistance at $B > B_{jump}$ to higher values.

Importantly, only the perpendicular component of the external magnetic field matters for switching the conductivity state. In the inset to Fig. 3d, the magnetic field value where the resistance drops ($B_{jump}$) and its sample normal component ($B_z$) are plotted for several angles between the sample normal (0°) and the field direction. The switching occurs at the same value of $B_z$ for all angles.

We note that according to our results obtained for samples grown at various Mn fluxes, the relevant quantity influencing the resistance is not the absolute value of the hole density but rather the ratio of the hole density to Mn concentration. This is demonstrated in Fig. 3e where the relevant data for an "inverted" structure (sample B) with about three times larger Mn density and more than doubled 2D hole density $p$ is shown. The extraordinary magnetoresistance behaviour in this sample is virtually identical to that discussed above.

In order to interpret these findings we note that, as expected from theory of the Anderson-Mott localisation, when the carrier density is very low, the holes are strongly localised by the parent acceptors, independent of the strength of an external magnetic field. If, in turn, the hole density is sufficiently high, in particular greater than the Mn concentration, owing to the combined effect of many-body screening, large kinetic energy, and the extended character of wave functions corresponding to the upper Hubbard band of the acceptor states, the holes get delocalised, the effect visible in Figs. 2 and 3c. Most interesting is the intermediate range of carrier densities, in which the metallic-like conductivity is observed at high magnetic field, but a cross-over to the strongly localised regime occurs when the magnetic field decreases. This phenomenon, accounting for the celebrated colossal negative magnetoresistance[1-9], occurs if the spins do not show long range ferromagnetic order. As reviewed elsewhere[16], in such a case carrier localisation is enhanced at low magnetic fields by two effects: (i) spin disorder scattering on randomly oriented preformed ferromagnetic bubbles brought about by spatial fluctuations in the local density of carrier states near the Anderson-Mott localisation, and (ii) the decrease of the kinetic energy associated with the carrier redistribution over the two spin subbands. Obviously, the redistribution of holes between the relevant subbands depends strongly on the magnetisation direction[4,17], and is accompanied by an increasing contribution of light holes to the effective mass of carriers at the Fermi level, which enhances dramatically the field-induced hole delocalisation[4].

This scenario implies that no spontaneous long range ferromagnetic order develops in the InAs channel hosting the 2DHG in the relevant temperature range. This assumption seems to be true, as for the QWs containing weakly localised holes with the determined effective mass of $m^* = 0.16 m_o$ (see *Methods*) the expected Curie temperature is[14] $T_C < 20$ mK for the Mn content indicated by the SIMS measurements, $x < 10^{-4}$.

We now make evident that the high-resistance state, resistance jumps and hysteresis are due to the interplay of two bistabilities. First, following a recent theory[10], we address the consequences of bistability due to overheating of the hole gas in the region of high *electric* fields: owing to a strong temperature dependence of the resistance in weak magnetic fields and at fixed high bias voltage, the system of Mn spins and carriers is either in the overheated (low resistance) state or in the much less heated high resistance state. The starting point to show the significance of overheating at non-zero bias voltages is the heat balance equation,

$$U_{bias}^2/[R(T_h,B) L W] = F_S(T_h) - F_S(T_s), \qquad (1)$$

where $F_S(T)$ is the energy loss rate per unit of the QW surface $LW$, determined by the coupling to acoustic phonons at the hole and substrate temperatures $T_{h(s)}$, respectively[10]. The right hand side of Eq. 1 describes the energy flow from the hole bath to the phonon bath, provided by the Joule heating on the left hand side. Since the effects of the Mn spins on charge transport scale with magnetic susceptibility[4,5,9] and the resistance $R(T)$ varies exponentially[10] in this regime, (see inset of Fig. 2) we assume

$$R(T,B) = R_0 \exp(a[dB_{5/2}(T,B)/dB]/T)^\gamma, \qquad (2)$$

where $B_{5/2}(T,B)$ is the Brillouin function for $S = 5/2$; $R_0$, $a$, and $\gamma$ are parameters determined by fitting $R(T, B = 0)$ to the experimental zero-field resistance.

As shown in Fig. 3f, the model, developed with no further fitting parameters (see Supplementary Information), describes the presence of the resistance jumps, corresponding to the transition from the high-resistance state to the overheated low resistance state. However, as the potential drop and carrier cooling at the contacts is neglected[10], the range of $U_{bias}$, $T$ and $B$ where the jumps occur is lower than in the experiment.

While the above picture reproduces the jumps, occurring at different B-fields for up- and down-sweeps, the model fails to describe the hysteresis at low $B$, as the calculated resistance is symmetric with respect to $B$, in contrast to experiment. Actually, according to data in Fig. 3a and direct magnetisation measurements[18] the magnetic hystereses persist even for $U_{bias} \rightarrow 0$. At the same time, minor loops with magnetic field stopped for different time durations (Fig. 3e), point to a finite relaxation time of several minutes, implying the absence of ferromagnetic order.

We will now demonstrate that the presence of magnetic hysteresis without long range magnetic order follows from specific properties of our system. Unlike (Ga,Mn)As[19,20], compressively strained InAs QWs show a large energy separation between the heavy and the light hole subbands, which reduces an admixture of $j_z = \pm 1/2$ states to the wave functions of the holes residing in the ground state subband, $j_z = \pm 3/2$ (with the z-axis perpendicular to the QW plane). Accordingly, the heavy-hole intraband matrix elements of $j_x$ and $j_y$ are very small, so that the system is immune to an in-plane magnetic field, in agreement with the experimental findings shown in Fig.

3d. This means also the appearance of large anisotropy energy barriers for reversing the magnetisation direction of the preformed ferromagnetic bubbles. The corresponding relaxation time will grow exponentially with the number of holes contributing to the bubble, resulting in superparamagnetic-like metastabilities.

We expect the presence of a metastable behaviour in weak magnetic fields, even if the holes are bound by *individual* Mn acceptors. In this limit, the hole spin is coupled to the parent Mn acceptor by a strong antiferromagnetic exchange interaction[12], which in the present case assumes the Ising form $H = -\varepsilon_c j_z S_z$, where $\varepsilon_c = -(\beta/3)|\psi(0)|^2$. Here $\beta = -0.054$ eV nm$^3$ is the *p-d* exchange integral and $\psi(0)$ is the value of the acceptor envelope function at the Mn ion. The relaxation time $\tau_s$ between the two relevant heavy hole $j_z = \pm 3/2$ states is rather long, presumably in a millisecond to microsecond range[21,22]. A somewhat shorter relaxation time is expected for Mn spins in the relevant range of concentrations[23,24]. Under these conditions a direct hole spin relaxation is possible via a flip-flop, $j_z \rightarrow -j_z, -S_z \rightarrow S_z$ process. However, the corresponding rate is expected to be rather small, particularly at low temperatures, where it would involve particularly slow transitions between $\pm 5/2$ states of the ion in the orbital singlet state. Furthermore, the *p-d* coupling removes the degeneracy of the Mn spins states, which reduces the role of nuclear magnetic moments in the spin relaxation[24,25].

In this situation, spin relaxation of holes towards thermal equilibrium values of $<j_z>$ and, then, $<S_z>$, leading to the corresponding values of resistance, proceeds primarily *via* high energy intermediate states determined by the magnitude of $S_z$. The presence of the magnetic anisotropy barrier $E_a$ elongates the relaxation time $\tau_s$ of the hole spin by $\exp(E_a/k_B T)$. According to the quantitative model presented in Supplementary Information, the barrier vanishes in the magnetic field $B_c \approx \varepsilon_c S/(2\kappa\mu_B)$ and attains the value $E_a \approx \varepsilon_c Sj$ at $B = 0$ and $T \to 0$, where for the QW in question the Luttinger parameter is $\kappa = 7.53$ (ref. 26) and $\varepsilon_c$ varies between 0 and ~ 0.12 meV for the Mn acceptors at the QW edge and centre, respectively.

We find that for $\varepsilon_c = 0.12$ meV and $\tau_s = 1$ ms, $\tau$ reaches 100 s at 0.49 K, the time during which the magnetic field changes by ~0.4 T in our experiment. This explains the hysteretic behaviour and resistance relaxation in the way seen in millikelvin studies of molecular magnets[27,28] and individual rare earth ions[25]. In agreement with this evaluation, hystereses disappear above ~ 0.6 K. At the same time, the barrier is expected to vanish at $B_c = 0.34$ T, in accordance with the magnitudes of the apparent coercive fields at $T \to 0$. When the hole concentration increases, more and more holes occupy states with smaller values of $\varepsilon_c$ meaning that the barrier height, and thus the relaxation time and the width of the hystereses diminish, the effect visible in Fig. 3c. Moreover, the model invoking properties of individual acceptors rather than a collective action of many Mn ions explains why the observed behaviour scales with the ratio of the Mn to hole density and not with the Mn concentration.

In conclusion, our results demonstrate that the field-induced delocalisation of holes in Mn modulation-doped III-V QWs proceeds *via* an intermediate and novel metastable insulator phase. Within our model the jumps result from hole overheating whereas hysteresis stems from a large magnetic anisotropy of the heavy holes, coupled to the parent Mn acceptors by the strong *p-d* exchange interaction. The slow spin relaxation of individual bound holes revealed here, appealing from the viewpoint of quantum information processing and storing, should appear also in the case of single Mn acceptors residing in InAs quantum dots[29]. Interestingly, it may compete with spin quantum tunnelling at sufficiently low temperatures[25,27,28], provided that the decoherence rate of the complex will be smaller than the matrix element coupling the $S_z = \pm 5/2$ states.

**Methods**

The samples are patterned into standard L-shaped Hall bar geometries (1000 µm x 200 µm and 200 µm x 40 µm) employing optical lithography and wet chemical etching. Ohmic contacts are prepared by soldering alloyed InZn and annealed for 60 s at 30°C. For gate electric-field dependent measurements some samples are covered with a 130 nm thick insulating parylene film and a thin Ti/Au top gate electrode. The measurements are carried out either in a $^4$He bath cryostat or in a dilution refrigerator. Transport measurements in the low resistance range (high magnetic fields and/or temperatures above 1.5 K) are performed using standard low-frequency lock-in technique with operation currents of 100 nA. In the high resistance state (low temperatures and low magnetic fields), a constant voltage $U_{bias}$ was applied, the current through the sample was monitored, and the two-terminal resistance was calculated as $R_{2\text{-term}} = U_{bias}/I$. In the low B, low T regime, the magnetotransport depends on the sweep rate which was here set to 0.25 T/min.

The hole-density $p$ of the 2DHG was determined from classical Hall resistance and confirmed by the period of Shubnikov-de Haas oscillations. The characteristic value for the "normal" doped 2DHGs is $p = 4.3 \cdot 10^{11}$ cm$^{-2}$. The corresponding values for the "inverted" doped 2DHGs are $p = 4.4 \cdot 10^{11}$ cm$^{-2}$ and $11 \cdot 10^{11}$ cm$^{-2}$, for samples A and B, respectively. The values of the effective mass $m^* = 0.16 m_0$ determined from the cyclotron resonance measurements[30] for an "inverted" doped sample is larger than the one expected for the band-edge in-plane hole mass of the ground state subband in an infinitely deep InAs QW[31] but consistent with previous experimental studies[32].

**References**

1. Shapira, Y., Foner, S., Aggarwal, R. L. & Reed, T. B., EuO. II. Dependence of the insulator-metal transition on magnetic order. *Phys. Rev. B* **8**, 2316--326 (1973).

2. von Molnar, S., Briggs, A., Flouquet, J. & Remenyi, G. Electron localization in a magnetic semiconductor: $Gd_{3-x}v_xS_4$. *Phys. Rev. Lett.* **51**, 706--709 (1983).

3. Dagotto, E., Hotta, T. & Moreo, A. Colossal magnetoresistant materials: the key role of phase separation. *Phys. Reports* **344**, 1--153 (2001).

4. Wojtowicz, T., Dietl, T., Sawicki, M., Plesiewicz, W. & Jaroszyński, J. Metal-insulator transition in semimagnetic semiconductors. *Phys. Rev. Lett.* **56**, 2419--422 (1986).

5. Terry, I., Penney, T., von Molnár, S. & Becla, P. Low-temperature transport properties of $Cd_{0.91}Mn_{0.09}Te$:In and evidence of a magnetic hard gap in the density of states. *Phys. Rev. Lett.* **69**, 1800--803 (1992).

6. Oiwa, A., Katsumoto, S., Endo, A., Hirasawa, M. Iye, Y., Ohno, H., Matsukura, F., Shen, A. & Sugawara, Y. Giant negative magnetoresistance of (Ga,Mn)As/GaAs in the vicinity of a metal-insulator transition. *Phys. Status Solidi (b)* **205**, 167--171 (1998).

7. Smorchkova, I. P., Samarth, N., Kikkawa, J. M. & Awschalom, D. D. Spin transport and localization in a magnetic two-dimensional electron gas. *Phys. Rev. Lett.* **78**, 3571--574 (1997).

8. Smorchkova, I. P., Samarth, N., Kikkawa, J. M. & Awschalom, D. D. Giant magnetoresistance and quantum phase transitions in strongly localized magnetic two dimensional electron gases. *Phys. Rev. B* **58**, R4238--241 (1998).

9. Jaroszyński, J., Andrearczyk, T., Karczewski, G., Wróbel, J., Wojtowicz, T.,

Popović, D. & Dietl, T. Intermediate phase at the metal-insulator boundary in a magnetically doped two-dimensional electron system. *Phys. Rev. B* **76**, 045322 (2007).

10. Altshuler, B. L., Kravtsov, V. E., Lerner, I. V., Aleiner, I. L. Jumps in Current-Voltage Characteristics in Disordered Films. *Phys. Rev. Lett.* **102**, 176803 (2009).

11. Wurstbauer, U., Soda, M., Jakiela, R., Schuh, D., Weiss, D., Zweck, J. & Wegscheider, W. Coexistence of ferromagnetism and quantum Hall-effect in Mn modulation-doped two-dimensional hole system, *J. Cryst. Growth* **311**, 2160--162 (2009).

12. Ohno, H., Munekata, H., Penney, T., von Molnár, S. & Chang, L.L. Magnetotransport properties of *p*-type (In,Mn)As diluted magnetic III-V semiconductors. *Phys. Rev. Lett.* **68**, 2664-2667 (1992).

13. Bhattacharjee, A. K. & Benoit à la Guillaume, C., Model for the Mn acceptor in GaAs. *Solid State Commun.* **113**, 17--21 (2000).

14. Boukari, H. *et al.* Light and electric field control of ferromagnetism in magnetic quantum structures. *Phys. Rev. Lett.* **88,** 207204 (2002).

15. Wurstbauer, U. & Wegscheider, W. Magnetic ordering effects in a Mn-modulation-doped high mobility two-dimensional hole system. *Phys. Rev. B* **79**, 155444 (2009).

16. Dietl, T. Interplay between carrier localization and magnetism in diluted magnetic and ferromagnetic semiconductors. *J. Phys. Soc. Jpn.* **77**, 031005 (2008), and references therein.

17. Pappert, K., Schmidt, M.J., Humpfner, S., Rüster, C., Schott, G.M., Brunner, K., Gould, C., Schmidt, G. & Molenkamp, L.W. Magnetization-Switched Metal-Insulator Transition in a (Ga,Mn)As Tunnel Device. *Phys. Rev. Lett.* **97**, 186402 (2006).

18. Rupprecht, B., Krenner, W., Wurstbauer, U., Heyn, Ch., Windisch, T., Wilde, M. A., Wegscheider, W. & Grundler, D. Magnetism in a Mn modulation-doped InAs/InGaAs heterostructure with a two-dimensional hole system. *J. Appl. Phys.* **107**, 093711 (2010).

19. Sheu, B. L., Myers, R. C., Tang, J.-M., Samarth, N., Awschalom, D. D., Schiffer, P. & Flatté, M. E. Onset of Ferromagnetism in Low-Doped $Ga_{1-x}Mn_xAs$. *Phys. Rev. Lett.* **99**, 227205 (2007).

20. Myers, R. C., Mikkelsen, M. H., Tang, J-M., Gossard, A. C., Flatté, M. E. & Awschalom, D. D. Zero-field optical manipulation of magnetic ions in semiconductors. *Nature Mater.* **7**, 203 - 208 (2008).

21. Gerardot, B. D., Brunner, D., Dalgarno, P. A., Öhberg, P., Seidl, S., Kroner, M., Karrai, K., Stoltz, N. G., Petroff, P. M. & Warburton, R. J. Optical pumping of a single hole spin in a quantum dot. *Nature* **451**, 441-444 (2007).

22. Heiss, D., Schaeck, S., Huebl, H., Bichler, M., Abstreiter, G., Finley, J. J., Bulaev, D. V. & Loss, D. Observation of extremely slow hole spin relaxation in self-assembled quantum dots. *Phys. Rev. B* **76**, 241306(R) (2007).

23. Dietl, T., Peyla, P., Grieshaber, W. & Merle d'Aubigné, Y. Dynamics of spin organization in diluted magnetic semiconductors. *Phys. Rev. Lett.* **74**, 474--477 (1995).

24. Goryca, M., Ferrand, D., Kossacki, P., Nawrocki, M., Pacuski, W., Maślana, W., Gaj, J. A., Tatarenko, S. Cibert, J., Wojtowicz, T. & Karczewski, G. Magnetization Dynamics Down to a Zero Field in Dilute (Cd,Mn)Te Quantum Wells. *Phys. Rev. Lett.* **102**, 046408 (2009).

25. Giraud, R., Wernsdorfer, W., Tkachuk, A.M., Mailly, D. & Barbara, B. Nuclear Spin Driven Quantum Relaxation in $LiY_{0.998}Ho_{0.002}F_4$. *Phys. Rev. Lett.* **87**, 057203 (2001).

26. Sanders, G. D., Sun, Y., Kyrychenko, F. V., Stanton, C. J., Khodaparast, G. A., Zudov, M. A., Kono, J., Matsuda, Y. H., Miura, N. & Munekata, H. Electronic states and cyclotron resonance in *n*-type InMnAs. *Phys. Rev. B* **68**, 165205 (2003).

27. Friedman, J. R., Sarachik, M. P., Tejada, J. & R. Ziolo Macroscopic Measurement of Resonant Magnetization Tunneling in High-Spin Molecules. *Phys. Rev. Lett.* **76**, 3830-3833 (1996).

28. Thomas, L., Lionti, F., Ballou, R., Gatteschi, D., Sessoli, R. & Barbara, B. Macroscopic quantum tunnelling of magnetization in a single crystal of nanomagnets. *Nature* **383**, 145-147 (1996).

29. Kudelski, A., Lemaître, A., Miard, A. Voisin, P., Graham, T.C., Warburton, R.J. & Krebs, O. Optically Probing the Fine Structure of a Single Mn Atom in an InAs Quantum Dot. *Phys. Rev. Lett.* **99**, 247209 (2007).

30. Wurstbauer, U., Knott, K., Graf von Westarp, C., Mecking, N., Rachor, K., Heitmann, D., Wegscheider, W. & Hansen, W. Anomalous magnetotransport and cyclotron resonance of high mobility magnetic 2DHG in the quantum Hall regime. *Physica E* **42**, 1022-1025 (2010).

31. Suemune, I. Band-edge hole mass in strained-quantum-well structures. *Phys. Rev. B* **43**, 14099--106 (1991).

32. Oettinger, K., Wimbauer, Th., Drechsler, M., Meyer, B. K., Hardtdegen, H. & Luth, H. Dispersion relation, electron and hole effective masses in $In_xGa_{1-x}As$ single quantum wells. *J. Appl. Phys.* **79**, 1481-85 (1996).

**Acknowledgements**

UW, DW and WW acknowledge financial support from the Deutsche Forschungsgemeinschaft (DFG) via SFB 689, UW also acknowledges support from Hamburg Cluster of Excellence on *Nanospintronics*, whereas TD and CS acknowledge financial support by the Humboldt Foundation and EC project FunDMS (ERC Advanced Grant).

**Author contributions**

**Additional Information**

Supplementary Information is linked to the online version of the paper at www.nature.com/naturephysics. Reprints and permissions information is available at www.nature.com/reprints. Correspondence and requests for materials should be addressed to W.W.

**Figure Legends:**

**Figure 1: Results of four terminal magnetotransport measurements and sample architectures.** Longitudinal resistance (red) and Hall resistance (black) at 1.6 K for inverted (a) and normal (b) modulation doped QW structures showing well defined SdH oscillations and Hall plateaus. Inset in (a), (b), Schematic layer sequence of the inverted and normal Mn modulation doped QW structures. The QW consists of a strain relaxed InGaAs layer with an asymmetrically embedded strained InAs channel. (a), (b) Contrary to the normal doped QW structure (b), the low-field resistivity increases dramatically for the structure with an inverted doping layer (a) indicating strong hole localisation.

**Figure 2: Temperature dependence of longitudinal resistance of the inversely doped structure in high B-fields.** The longitudinal resistance $R_{xx}$ along the [-110] directions shows well developed SdH-oscillations with vanishing resistance at filling factor $\nu = 1$. At low B the system undergoes a quantum Hall insulator transition with a dramatic increase of $R_{xx}$ for all temperatures. The Hall resistance is shown for $T = 30$ mK. Inset: Temperature dependence of the zero field resistance ($U_{bias} = 0.5$ V) at various gate voltages demonstrating the tunability from strong to weakly localized with increasing carrier density.

**Figure 3: Two terminal resistance of the inverted structure – sample A (a-d); sample B (e); computed (f).** (a) Magnetoresistance showing temperature-dependent hystereses and abrupt resistance drops for $U_{bias} = 0.01$ V (a) and $U_{bias} = 0.5$ V (b). $R_{max} = 10^{10}$ Ω (a) and $R_{max} = 10^{12}$ Ω (b) denotes an upper limit of the experimental set-up. Sweep directions are indicated by arrows. (c) Magnetoresistance at different

carrier densities varied by the top-gate voltage at $T$ = 170 mK. (d) Current–voltage characteristics at $T$ = 40 mK for selected values of perpendicular magnetic field (sweep direction marked by an arrow) showing highly non-linear and linear behaviour at low and high magnetic fields, respectively. The inset shows the magnitude of the total magnetic field corresponding to the abrupt resistance decrease, $B_{jump}$ (black), and the normal component of $B_z$ (red) as function of the angle between the magnetic field direction and the film normal at 40 mK. (e) Minor loops for sample B at 400 mK, documenting enhanced relaxation times. B-field sweeps in upward direction were stopped at $B$ = 0.31 T. Then, after different waiting times indicated in the figure, the magnetic field was swept down. Open black squares show a full up-sweep (after complete relaxation). (f) Computed resistance as a function of the magnetic field at various substrate temperatures for $L$ = 0.9 mm and $U_{bias}$ = 0.11 V for sample A. Arrows mark positions of resistance jumps.

*Figure 1*

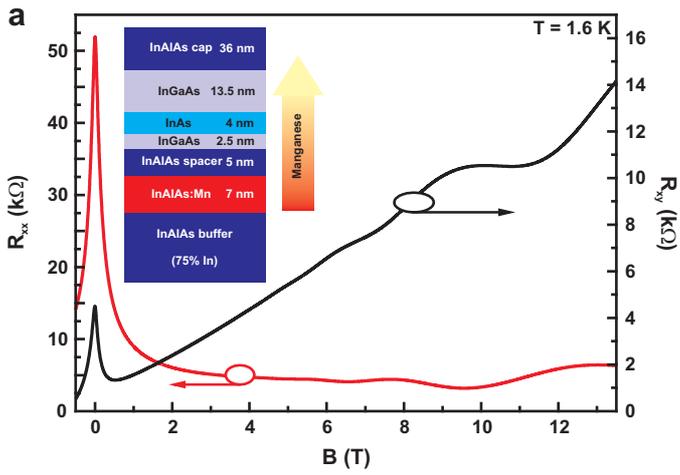

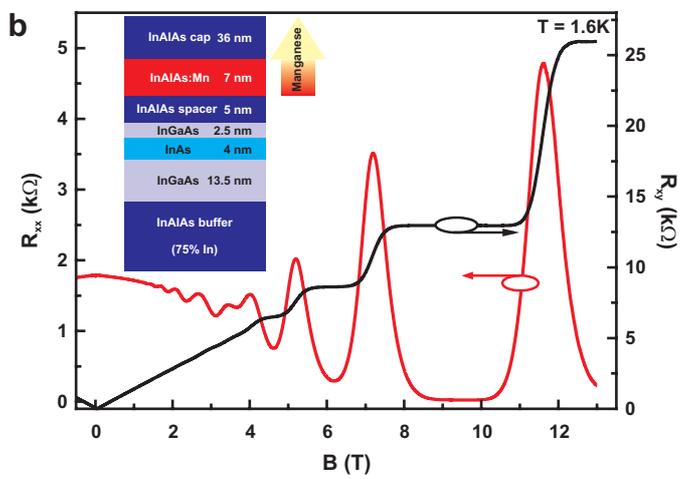

*Figure 2*

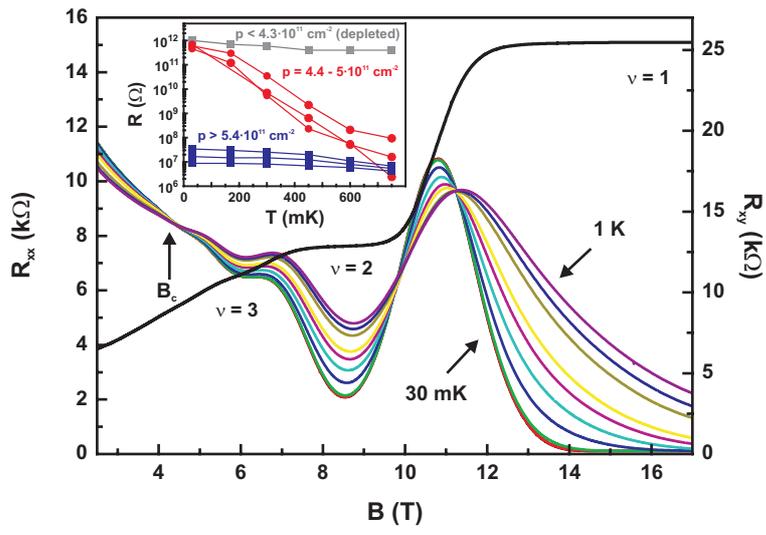

*Figure 3*

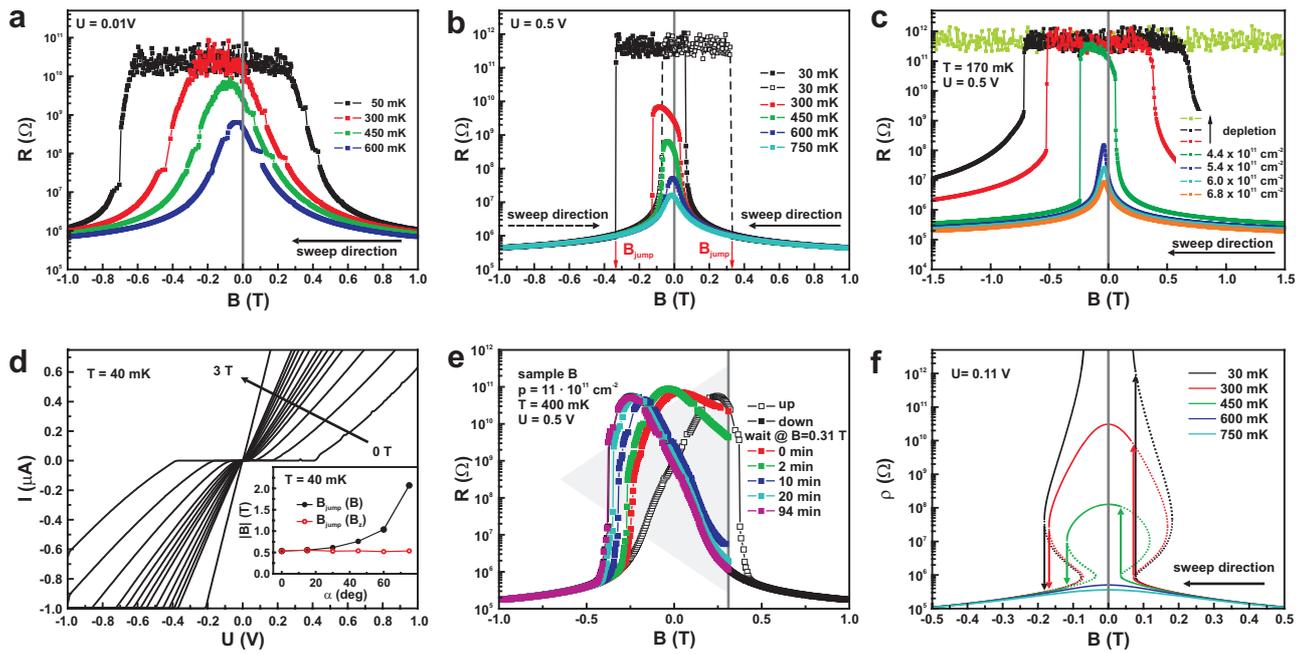